\documentclass[]{spie}  

\usepackage{amsmath,amsfonts,amssymb}
\usepackage{graphicx}
\usepackage[colorlinks=true, allcolors=blue]{hyperref}

\usepackage[dvipsnames]{xcolor}
\usepackage[normalem]{ulem}

\title{Advances in entanglement-based QKD for space applications} 

\author[a]{Sebastian Ecker}
\author[a,b]{Johannes Pseiner}
\author[b]{Jorge Piris}
\author[c]{Martin Bohmann}

\affil[a]{Institute for Quantum Optics and Quantum Information (IQOQI), Boltzmanngasse 3, 1090 Vienna, Austria}
\affil[b]{European Space Research and Technology Centre (ESTEC) - European Space Agency (ESA), Keplerlaan 1, Postbus 299, 2200 AG Noordwijk, Netherlands}
\affil[c]{Quantum Technology Laboratories GmbH (qtlabs), Clemens-Holzmeisterstr-Str. 6/6, 1100 Vienna, Austria}

\authorinfo{Further author information: (Send correspondence to S.E. or M.B.) \\S.E.: E-mail: sebastian.ecker@oeaw.ac.at \\M.B.: E-mail: martin.bohmann@qtlabs.at}

\begin{document}
\maketitle

\begin{abstract}
Quantum key distribution (QKD) enables tap-proof exchange of cryptographic keys guaranteed by the very laws of physics. One of the last remaining roadblocks on the way towards widespread deployment of QKD is the high loss experienced during terrestrial distribution of photons, which limits the distance between the communicating parties.  A viable solution to this problem is to avoid the terrestrial distribution of photons via optical fibers altogether and instead transmit them via satellite links, where the loss is dominated by diffraction instead of absorption and scattering. First dedicated satellite missions have demonstrated the feasibility of this approach, albeit with relatively low secure key rates. In order for QKD to become commercially viable, the design of future satellite missions must be focused on achieving higher key rates at lower system costs. Current satellite missions are already operating at almost optimal system parameters, which leaves little room for enhancing the key rates with currently deployed technology. Instead, fundamentally new techniques are required to drastically reduce the costs per secret bit shared between two distant parties. Entanglement-based protocols provide the highest level of security and offer several pathways for increasing the key rate by exploiting the underlying quantum correlations. In this contribution, we review the most relevant advances in entanglement-based QKD which are implementable over free-space links and thus enable distribution of secure keys from orbit. The development of satellite missions is notoriously lengthy. Possible candidates for a new generation of quantum payloads should therefore be scrutinized as early as possible in order to advance the development of quantum technologies for space applications.
\end{abstract}


\section{INTRODUCTION}

    In our modern information-technology driven world, secure communication is getting ever more important.
    To secure the privacy of communication channels, cryptographic tools are used.
    Encryption systems have been constantly further developed, driven by the competition with code breakers, who always find more elaborate ways to challenge secure communication. 
    Today, the computational power of supercomputers and even more so the rise of quantum computers pose a substantial threat to the security and privacy of our communication channels which has a dramatic impact on our private, political, and economical life.
    This threat stems from the fact that modern encryption is based on mathematical problems which are hard to solve by classical computers, but can be efficiently solved on a quantum computer.
    
    One possibility of overcoming this quantum threat also lies in the quantum realm; namely quantum communication and more particular quantum key distribution (QKD).
    QKD is an approach to exchange a secure key for encryption between two or more users guaranteeing security by the laws of quantum physics.
    An extensive introduction and overview of the field can be found, for example, in the following review articles \cite{Gisin2002,Gisin2007,Scarani2009,Xu2020}.
    Here, we will just briefly recall the main ideas and concepts.
    Central to QKD is the so-called no-cloning theorem that states that an arbitrary quantum state, and therefore the quantum information it is carrying, cannot be simply copied, which is the key ingredient for the security of QKD.
    Due to their transmission properties and minimal interaction with the environment, quantum states of light are usually used to encode and distribute quantum information based on which it is possible to extract a secure key for encryption afterwards.
    In QKD, there are two main branches, namely the continuous-variable (CV) and discrete-variable (DV) regime.
    In the former, quantum information is encoded in the field properties of an optical mode while in the latter one properties and correlations of individual photons are used. 
    In the DV regime the most common approaches are the prepare and measure protocols based on so-called BB84 protocol \cite{Bennett1984} and entanglement-based approaches which exploit quantum entanglement between two photons \cite{Ekert1991,Bennett1992}.
    As the name already suggests, prepare and measure protocols need an active encoding of the information onto the quantum state of a single photon or weak coherent states if one uses the decoy-state protocol\cite{Lo2005}. 
    This active encoding demands that one communication partner (usually called Alice) possesses the sending device, that the device is trusted, and that one has access to true random numbers for the active encoding.
    On the other hand, entanglement-based schemes are based on the creation of two entangled photons through a physical process and the subsequent distribution of these photons to the communication partners Alice and Bob.
    This approach does not require that one of the communication partners possesses the source of entangled photons nor is it necessary to trust the device as the quantum correlations between the two photons measured by the communication partners cannot be emulated or faked by a malicious adversary.   
    For BB84 and entanglement-based protocols one has to choose the degree of freedom (physical property) in which one wants to encode and measure the quantum information that can be, for example, polarization, time, or orbital angular momentum.
    Note that there exist also other QKD schemes such as measurement-device independent or twin-field QKD; again, see Refs. \citenum{Gisin2002,Gisin2007,Scarani2009,Xu2020} for an overview. 
    
    Let us briefly recall some further fundamentals of entangled-photon sources.
    In principle, there are several physical processes that allow to generate entangled photons for QKD such as exciton decay in quantum-dots \cite{Schimpf2021} or nonlinear optical processes like spontaneous parametric down-conversion (SPDC) (see, e.g. Ref.~\citenum{Couteau2018}) or four-wave mixing \cite{McMillan2013}.
    SPDC is the most advanced and most broadly used technique for implementing entanglement-based QKD.
    Its basic working principle is the conversion of pump photons in a material with second-order nonlinearity into two (down-converted) photons that are entangled.
    Depending on the design of the SPDC source, the generated photons are entangled in different degrees of freedom and can even feature simultaneous entanglement -- so-called hyperentanglement -- in several degrees of freedom \cite{Kwiat1997}.
    Notably, high-dimensional degrees of freedom provide a larger information content per sent photon \cite{Barreiro2008} and show better resistance against noise \cite{Ecker2019}.
    An overview on different designs of entangled photon sources (EPS) can be found in Ref.~ \citenum{anwar2021entangled}.
    Among others, important parameters of SPDC source are: the SPDC emission spectrum, the source's heralding efficiency, the brightness describing the average number of entangled photon pairs per second (and per pump power and spectral width), and deviations from the ideal (maximally entangled) target state.
    A definition of these properties and their influence on the performance of QKD protocols can be found in Ref. \citenum{Neumann2021}.
    It is also important to mention that entangled photons cannot only be used for QKD applications but they will be an essential building block of a future quantum internet.
    
    The fundamental principle which provides security in QKD systems  also limits the implementation, as quantum information cannot be simply amplified resulting in the fact that losses (attenuation) play a crucial role in QKD.
    Therefore, the implementation of QKD systems is practically limited to a few hundred kilometers in optical fibers \cite{Wengerowsky2019,Neumann2022}, because the losses in fiber systems scale exponentially with the communication distance.
    A possible way to overcome this problem is to use satellite-to-ground links, which do not scale exponentially with the distance between the communication parties and therefore even allow for global quantum communication.
    An overview on free-space satellite QKD implementations is provided in Section \ref{sec:motivation}.
    The implementation of free-space QKD links, however, needs to deal with the influence of the turbulent atmosphere \cite{Vasylyev2016,Vasylyev2019} that can have a detrimental effect on the transmitted quantum correlations \cite{Bohmann2016,Bohmann2017,Hofmann2019}.
    Additionally, the losses in satellite-based QKD are still high which limits the attainable key rate.
    Therefore, it is important to develop strategies to optimize and maximize the achievable secure key rate in satellite QKD systems.

    In this paper, we present potential advances in future entanglement-based QKD systems involving satellites, including their demands and challenges. 
    The discussed approaches cover the exploitation of physical properties as well as technological advances.
    We will elaborate on the development of high-performance and integrated entangled photon sources.
    Furthermore, we will elaborate how wavelength-multiplexing and high-dimensional QKD can be beneficial. 
    Finally, we also discuss the potential of adaptive optics and space-based quantum repeaters for quantum communication.
    
    This paper is organized as follows.
    In Section \ref{sec:motivation}, we provide a motivation and the state of the art for entanglement-based QKD from space.
    In Section \ref{sec:advances}, we list and discuss different pathways for advances in space-based QKD.
    We finish with a conclusion and discussion in Section \ref{sec:conclusion}.





\section{MOTIVATION AND STATE OF THE ART}
\label{sec:motivation}
Prior to any satellite launch, the idea of using long-distance optical free-space links for QKD was successfully put to the test over terrestrial links~\cite{SchmittManderbach2007,Ursin2007,Fedrizzi2009}. 
These studies demonstrated the feasibility of satellite-based QKD by comparing the atmospheric turbulences on a horizontal free-space link with the turbulences experienced over a vertical free-space link~\cite{Ursin2007}.
In 2016, the first satellite with a dedicated quantum payload, named \textit{Micius}, was successfully launched. 
This satellite launch was part of the QUESS mission~\cite{khan2018satellite}, spearheaded by the Chinese Academy of Sciences, which demonstrated several quantum communication protocols via satellite links.
Specifically, several QKD configurations, such as satellite-relayed QKD~\cite{Liao2018}, satellite-to-ground QKD~\cite{Liao2017,Yin2017} and entanglement-based QKD via a dual downlink~\cite{Yin2020}, have been successfully demonstrated. 
Several other missions have deployed dedicated satellite transmitter payloads which served as a driver for the development of robust quantum technologies~\cite{Steinlechner2016,Tang2016a,Tang2016b,Takenaka2017}.

The preferred photonic property for quantum state encoding over free-space links is the polarization domain due to its robustness in free-space propagation, the availability of high-quality polarization-encoded photon sources and the availability of efficient and compact receiving modules. 
Most advances discussed in the following will therefore rely on the polarization domain for key generation.

Another important design choice concerns the type of QKD protocol. 
While the first QKD protocol was of the prepare-and-measure type~\cite{Bennett2014}, where Alice prepares the quantum states and Bob receives them, entanglement-based protocols~\cite{Ekert1991,Bennett1992}, where both Alice and Bob receive photons from an entangled photon pair source, have already been successfully demonstrated in space~\cite{Yin2020}.
Entanglement-based QKD protocols are advantageous for space applications for two reasons. 
In contrast to prepare-and-measure protocols, the party owning the source of entangled photon pairs can be malicious, since the communicating parties can infer any eavesdropping attempt from the correlations between their measurement outcomes. 
This is particularly relevant for space-based quantum cryptography, since the satellite operator hosting the entangled photon pair source can in principle be malicious, which is essential, since trust in the cryptographic devices is not a good selling point in QKD implementations.  
Apart from that, entanglement itself is a valuable resource in quantum information processing~\cite{Horodecki2009}, and the long-distance distribution of entanglement is therefore crucial for future quantum technological applications in space.

The secure key rate distributed via a dual downlink from Micius between two parties on ground separated by 1120 km was 0.12 bits/s~\cite{Yin2020}. 
While this is impressive for a first in-orbit demonstration, the commercial success of QKD is directly linked to the costs per secret bit. 
Importantly, the hardware employed on the satellite and on the ground stations is already operating close to the optimum~\cite{Ecker2021}, which leaves little room for further optimisation. 
In order for QKD to be commercially viable, fundamentally new methods and techniques are required to increase the secure key rate and lower the costs per secret bit.
Fortunately, the quantum correlations which are inherent to entanglement-based QKD protocols open up several pathways for increasing the secure key rate with reasonable technological overhead.  

\newpage
\section{ADVANCES IN SPACE-BASED QKD}
\label{sec:advances}
In each of the following subsections, a method towards advancing satellite-based QKD is introduced. 
While all of the methods are compatible with satellite links, the same methods can also be employed to increase the key rate in fiber-based QKD. 
All of the presented advances have a technology readiness level of 4-7, which means they have been demonstrated in laboratory environments and some even have seen demonstrations over terrestrial free-space links. 
The introduced advances comply with the following key requirements:
\begin{itemize}
\item Increase the secure key rate
\item Reduce size, weight, power (SWaP) and complexity of quantum payloads
\item Decrease the costs per secret bit
\end{itemize}
In the Discussion section as well as in each of the following subsections we assess and rank these advances according to their space suitability, their deployment timeline and their key rate potential. 

\subsection{High-performance entangled photon-pair sources}
Entangled photon pair sources play a crucial role in the overall performance of a QKD system, and they have been the bottleneck in the achievable key rate for some time.
The performance of entangled photon-pair sources for QKD can be captured by three parameters:
\begin{itemize}
\item \textbf{Brightness}:
Photon pair rate per unit of pump power
\item \textbf{Fidelity}: Closeness of the produced quantum state to a maximally entangled quantum state 
\item \textbf{Heralding efficiency}: Probability of a photon-pair detection conditioned on a single-photon detection
\end{itemize}
However, only one of the three parameters is relevant for state-of-the-art entanglement sources. 
While the fidelity of the source has a huge impact on the achievable secure key rate, most polarization-entangled photon pair sources based on SPDC are characterized by fidelities in excess of 99\%~\cite{MeyerScott2018}, which means they emit almost perfect quantum states.
The potential of achieving slightly higher key rates by investing in this remaining 1\% for achieving 100\% fidelity is currently not worth the effort.
Optimizations on the heralding efficiency are also insignificant, since slightly higher heralding efficiencies are overshadowed by high optical losses in satellite links. 

\begin{figure} [ht]
\begin{center}
\includegraphics[height=5cm]{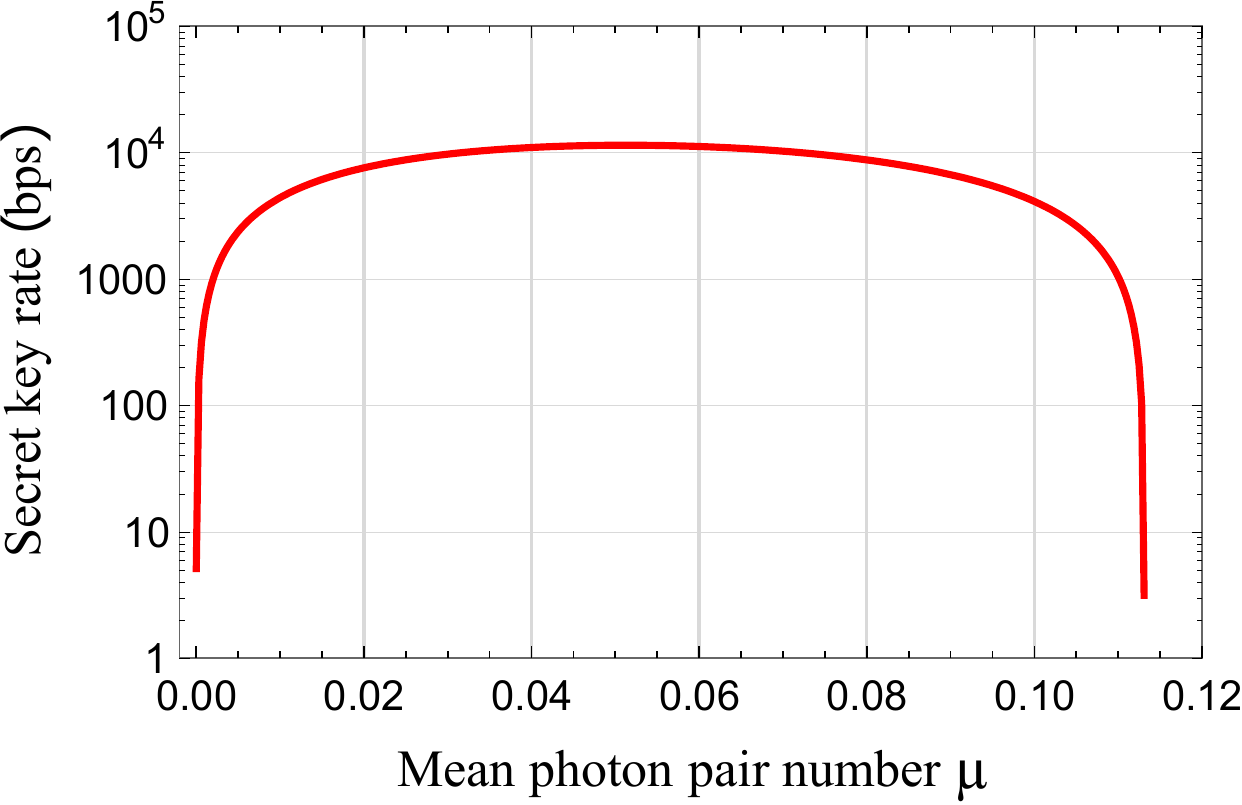}
\end{center}
\caption[example] 
{\label{fig:acccoinc} 
Secret key rate as a function of mean photon pair number per coincidence window. Increasing the mean photon pair number eventually results in a decrease in secret key rate due to an increasing probability of detecting uncorrelated photon pairs, so-called accidental coincidences.}
\end{figure}

The brightness of photon pair sources is the one parameter which can make a difference in secure key rate. 
Naively, one might expect that an increase in brightness corresponds to a proportional increase in secure key rate. 
Due to so-called \textit{accidental coincidences}, this is however not always the case. 
Pairs of photons produced in SPDC are highly correlated on a ps timescale. 
They are identified by correlating the detection times of Alice's and Bob's detection events. 
Whenever a photon detection at Alice coincides with a photon detection at Bob within a certain coincidence window, a photon pair is identified.
With increasing source brightness and background noise levels, detection events are mistakenly identified as a photon pair. 
These accidental coincidences cannot be distinguished from genuine photon-pair detections, which results in a decreased fidelity and therefore potentially results in a decreased key rate (see Figure~\ref{fig:acccoinc}).
These fake identifications become more frequent if the pair rate is increased beyond the timing resolution of the detection system. 
This necessitates an adaptation of the pair rate to the available timing resolution of the detection system and the chosen coincidence window. 
While single-photon avalanche diodes (SPADs) have a typical timing resolution in the order of 100 ps, another widespread detector technology, superconducting nanowire single-photon detectors (SNSPDs)~\cite{Natarajan2012}, have typical timing resolutions in the order of 10 ps and therefore enable significantly smaller coincidence windows. 
The design of QKD systems therefore requires careful fine-tuning of the transmitter and receiver specifications.

While this is true for static quantum links, dynamically changing quantum links require adaptive measures for efficient operation even more. 
For example, in LEO satellite links the optical loss varies drastically depending on the elevation angle, weather conditions and cloudage.
Therefore, adapting the photon pair rate via the pump power of the SPDC process is essential for optimizing non-static satellite links. 
This has recently been demonstrated over a 143-km-long terrestrial free-space link between two Canary islands, where one photon is measured locally and its entangled partner photon is sent over the free-space link~\cite{Ecker2021}. 
Depending on the attenuation of the free-space link, the optimal key rate was found at a different source brightness. 

Crucially, the optimal photon pair rates can be achieved with state-of-the-art entanglement sources. 
Even the entanglement source on the Micius satellite~\cite{Yin2017} was in principle able to operate close to the optimal rate~\cite{Ecker2021}.
Thus, the brightness of state-of-the-art entanglement sources is sufficient for optimal operation of satellite-links for QKD. 
This is true for detection time resolutions in the order of hundreds of ps, which are typical for SPADs used for free-space applications. 
However, even commercially available SNSPDs with time resolutions below 50 ps can be used in QKD systems while still operating at the optimal photon pair rate~\cite{Neumann2021}.

Entangled photon pair sources might again be challenged to achieve ever-higher pair rates once SNSPDs become commercially available which have timing resolutions in the order of a few ps~\cite{Korzh2020}.  
Another leap in brightness can be expected if wavelength multiplexing is widely adopted (see section~\ref{sec:multiplex}). 
While achieving such high pair rates is a challenge in itself, another challenge must be overcome first – namely the coupling of a free-space beam to the active detector area of a SNSPD. 
SNSPDs cannot be straightforwardly employed for free-space applications, since they currently only interface with single-mode fibers. 
In order to make SNSPDs free-space compatible, the multi-mode beam could be either coupled into single-mode fibers with adaptive optics (see section~\ref{sec:adaptopt}) or directly impinge on the nanowire detector through a vacuum viewport, as demonstrated in Ref.~\citenum{Mueller2021}.

\subsection{Integrated photonic entanglement sources} 
Photonic integrated circuits (PIC) are miniaturized versions of table-top bulk optical setups. 
They are primarily used in the telecommunication industry, for biomedical applications and for photonic quantum information processing.
An increasing number of optical operations can be hosted on PICs and the opto-electronic integration has advanced to the point where lasers, photon sources and detectors can be implemented on a single chip. 
Here, we are interested in PICs as integrated photonic entanglement sources for space deployment. 
The benefits of switching from bulk-optic sources to integrated sources are manifold: 
\begin{itemize}
    \item Drastically reduced SWaP
    \item Non-linearities for photon pair generation are inherent to the material platforms
    \item Robustness (no misalignment during launch) and phase stability
    \item Laser integration via electrical injection~\cite{Boitier2014}
    \item Scalability in production (fabrication facilities already exist for some platforms\footnote{Silica-on-silicon integrated optics is compatible with standard CMOS fabrication})
    \item Multiple sources on a single chip - avoiding a single point of failure and achieving parallelization
\end{itemize}
There are two processes, inherent to materials used for PICs, which enable photon pair production. 
Spontaneous four-wave mixing (SFWM) is a third-order nonlinearity, which involves two pump fields which are converted into a two-photon state.
It is present in materials such as silicon (Si), silica (SiO\textsubscript{2}), silicon nitride (Si\textsubscript{3}N\textsubscript{4}) or silicon oxy-nitride (SiO\textsubscript{x}N\textsubscript{y}). 
SPDC, a second-order nonlinear process, is present in III-V semiconductors such as gallium arsenide (GaAs) or aluminium gallium arsenide (AlGaAs), or in waveguides based on lithium niobate (LN).

Although SPDC requires careful phase-matching of vastly different wavelengths, for bright entanglement sources it is preferable to SFWM for a number of reasons. 
Firstly, similar wavelengths of the strong pump and photon pairs in SFWM render pump rejection a major challenge, while this is relatively easy for SPDC. 
Secondly, quasi phase-matching in SPDC enables effective nonlinearities which are much larger than in SFWM. 
Thirdly, an additional problem encountered in Si waveguides are two-photon and three-photon absorption, which limit the heralding efficiency and brightness of the source.

While integrated photonic sources for entanglement in the energy and time domain have been developed~\cite{Sciara2021}, polarization-entanglement production is particularly relevant, since polarization-entangled sources require phase-stable overlap of two SPDC processes. 
The most promising platforms for polarization-entangled integrated photon pair sources which meet the requirements for long-distance QKD are periodically-poled waveguide sources~\cite{Sansoni2017,Atzeni2018} and monolithic III-V semiconductor sources~\cite{Boucher2014,Zhukovsky2012,Valls2013,Horn2013}, which are summarized in Table \ref{tab:integrsources}.

\begin{table}[ht]
\caption{Specifications of two promising candidates for integrated polarization-entangled photon pair sources for long-distance QKD.} 
\label{tab:integrsources}
\begin{center}       
\begin{tabular}{|c||c|c|} 
\hline
\rule[-1ex]{0pt}{3.5ex}   & Periodically-poled waveguide sources & Monolithic III-V semiconductor sources  \\
\hline\hline
\rule[-1ex]{0pt}{3.5ex}  Material platform & periodically-poled LN (ppLN) & GaAs, AlGaAs   \\
\hline
\rule[-1ex]{0pt}{3.5ex}  Material assembly & hybrid glass-LN & ridge Bragg-reflection waveguide  \\
\hline
\rule[-1ex]{0pt}{3.5ex}  Phase-matching & quasi phase-matching & Bragg modes and total internal reflection  \\
\hline
\rule[-1ex]{0pt}{3.5ex}  Brightness & $\sim 5\cdot10^6$ pairs/s/nm/mW & $\sim 10^6$ pairs/s/nm/mW  \\
\hline
\rule[-1ex]{0pt}{3.5ex}  Footprint & $\sim$ cm & $\sim$ mm  \\
\hline
\end{tabular}
\end{center}
\end{table}
Both of these integrated photon pair sources promise brightness values comparable to the best bulk optical implementations, while their footprint is orders of magnitude smaller. 
While the state fidelities of integrated sources have not reached the values typical for bulk optical sources, one must keep in mind that bulk optical sources have undergone many years of research, while integrated photon pair sources still have a huge potential for improvements.

For space deployment of integrated photonic sources, several additional aspects must be taken into account.
The produced photon pairs must be efficiently coupled into single-mode optical fibers both for spatial-mode filtering and for guiding the photons to the transmitter telescopes. 
Several strategies for coupling light out of PICs exist, which can be coarsely divided into edge couplers and vertical couplers (e.g. grating couplers) \cite{Marchetti2019}. 
While vertical couplers enable vertical access to PICs and have relaxed positioning tolerances, often times they are polarization and wavelength sensitive.
On the other hand, edge couplers are hardly polarization sensitive and lead to higher coupling efficiencies, while the positioning between the fiber and the PIC is more delicate. 
Alternatively, the entangled photons can be directly launched without any intermediate fiber coupling.
This requires precise pre-alignment of the PIC relative to the collimation lens or transmitter telescope.
Another important aspect is the packaging and assembly of the integrated photon pair source, which is particularly crucial since it must withstand launch conditions. 
In addition to guaranteeing permanent coupling between the fibers and the PIC, electrical connectivity to phase-shifters or to electrically-injected sources must be ensured after packaging. 
Crucially, the integrated photon pair source must be temperature-stabilized since the phase matching is highly temperature-dependent. 

Apart from transmitter and receiver optics, integrated photonics is an obvious contender for replacing bulk optical components in space, and will play a crucial role both for classical optical communication and quantum communication.

\subsection{Wavelength-multiplexed QKD} %
\label{sec:multiplex}
Frequency-division multiplexing is the basis for modern high-rate optical telecommunication networks.
In QKD, wavelength multiplexing serves a similar purpose, leading to a substantial increase in secure key rate. 

\begin{figure} [t]
\begin{center}
\includegraphics[width=0.7\textwidth]{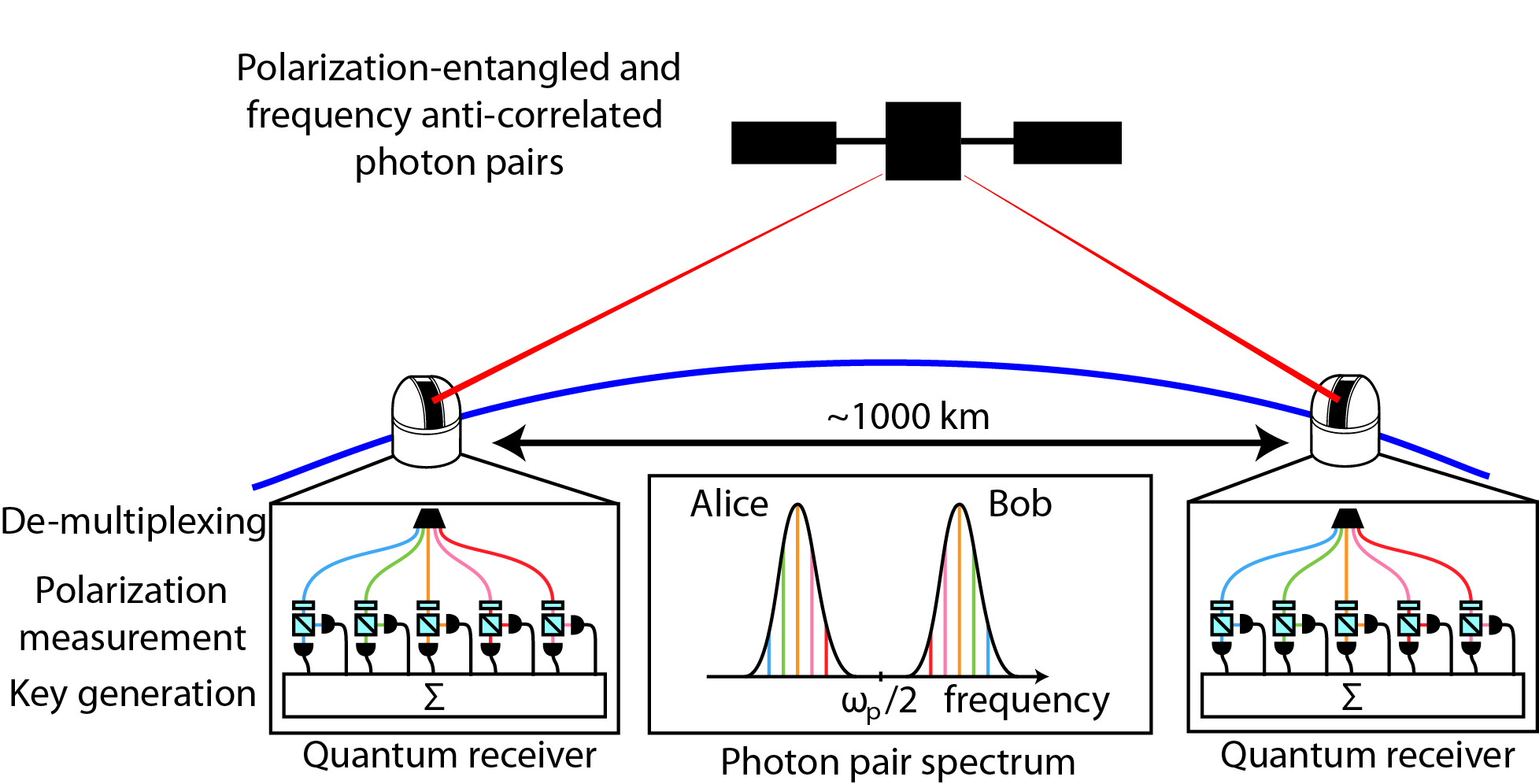}
\end{center}
\caption[example] 
{\label{fig:multiplexing} 
Illustration of a wavelength-multiplexed QKD system for space applications. A satellite establishes a dual downlink to Alice's and Bob's ground stations with polarization-entangled and frequency anti-correlated photon pairs produced in SPDC. Alice and Bob perform frequency-resolved measurements of the polarization and extract a secure key from each frequency channel individually. The inset illustrates a typical photon pair spectrum emitted in SPDC. The two envelopes correspond to the intensity spectra of Alice's and Bob's photon. Due to energy conservation, all frequency channels which are symmetric about $\hbar \omega_\text{p}/2$ are anticorrelated and can be used for wavelength-division multiplexing.}
\end{figure}

In frequency-division multiplexing, each frequency band is used as a separate signal carrier. 
After multiplexing, all of the frequency channels are transmitted along the same optical fiber and de-multiplexed at the receiver. 
For photon pairs produced in SPDC, both the signal modulation as well as the multiplexing step are redundant.
The signal modulation is unnecessary, since the polarization-entangled two-photon state is identical for each produced photon pair and the key is not produced until Alice and Bob perform measurements on their photons.  
Due to energy conservation in the SPDC process, the energy of the pump photon $\hbar \omega_\text{p}$ is equal to the combined energy of the signal $\hbar \omega_\text{s}$ and idler $\hbar \omega_\text{i}$ photons, resulting in $\hbar \omega_\text{p} = \hbar \omega_\text{s} + \hbar \omega_\text{i}$. 
For a spectrally narrowband pump laser, the frequencies $\omega_\text{s}$ and $\omega_\text{s}$ are therefore strictly anticorrelated (see inset Figure~\ref{fig:multiplexing}), which means frequency multiplexing in opposite frequency bands with respect to $\hbar \omega_\text{p}/2$ is inherent to the SPDC process. 

Importantly, the rate increase is not caused by adding wavelength channels to the spectrum.
An increased secure key rate can also be observed if instead of detecting the entire spectrum with a single detector, the spectrum is frequency-resolved in N channel pairs and detected by N detectors both at Alice and Bob.
While this might seem counterintuitive, the key rate increase is not rooted in an increased photon pair rate, but exclusively in preventing accidental coincidences, which is also why this method is effective close to the optimal source brightness, where accidental coincidences already start to contribute.
For N frequency channels, a single photon-pair source effectively acts as N separate sources, where each source operates at different, anticorrelated frequency pairs. 
If the source was operated at the optimal brightness before multiplexing, after multiplexing, the brightness of the photon pair source can be N-fold increased, which leads to an N-fold increase of the total secure key rate. 
This simplified example does not take into account misfirings of the detectors, so-called detector dark counts, which can have a substantial contribution to the accidental coincidences in high-loss scenarios.

Wavelength multiplexing can be realized through a number of chromatic dispersive effects. 
For fiber-based telecommunication, wavelength division multiplexing is widely deployed and standards such as dense wavelength-division multiplexing (DWDM) form the backbones of the internet. 
DWDM hosts up to 160 channels with channel spacings down to 12.5 GHz in the telecommunication range. 
While the technology is readily available and cheap, it is not directly applicable to free-space communication, since it is based on fiber Bragg gratings in single-mode waveguides, and therefore requires adaptive optics (see Sec.~\ref{sec:adaptopt}).
In order to spectrally decompose a free-space beam, the most commonly used devices are prisms and gratings. 
Reflective gratings can be realized via periodic structures on the surface of a material which lead to a position-dependent phase-change of the light. 
Another type of grating utilizes periodic refractive index changes within a bulk prism. 
These so-called volume holographic gratings (VHGs) can be either reflective or transmittive and they provide narrow channel spacings between 0.01 and 0.1 nm. 
By stacking several VHGs, a highly efficient wavelength de-multiplexer can be realized, where each VHG reflects a narrow band of the spectrum at a specific angle (the Bragg angle) and transmits the rest of the spectrum~\cite{Pseiner2021}. 
In this way, many wavelength channels can be accessed in a scalable way.

\begin{figure} [t]
\begin{center}
\includegraphics[width=\textwidth]{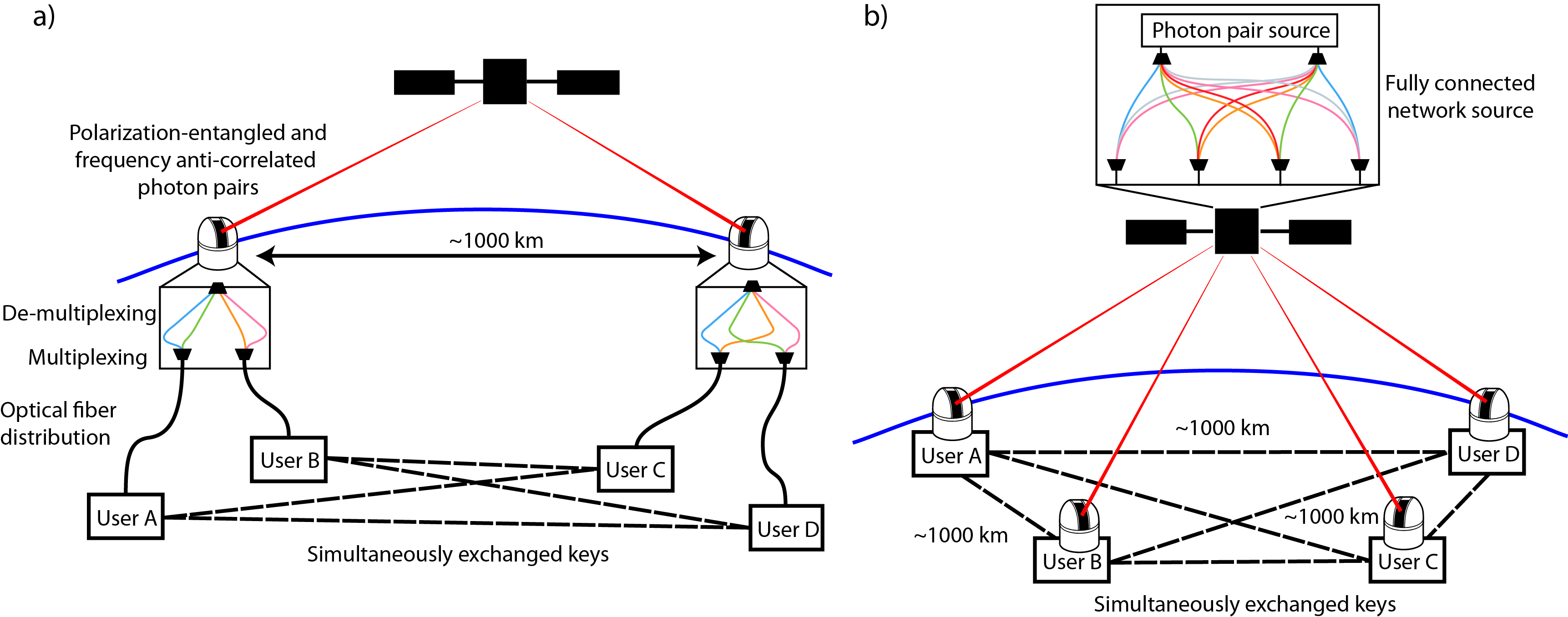}
\end{center}
\caption[example] 
{\label{fig:network_ground} 
Two possible configurations for satellite-based QKD exploiting wavelength-division networking. a) A dual downlink from a satellite distributes polarization-entangled and frequency anti-correlated photon pairs between two ground stations. The ground stations frequency-resolve the received photons and selectively multiplex them into four single-mode fibers. These fibers are then guided to the end users, which can simultaneously exchange keys with other end users in remote cities. b) The frequency anti-correlations of a polarization-entangled photon pair source on a satellite are exploited to construct a fully connected network on the satellite~\cite{Wengerowsky2018}. Then, the four resulting source outputs are distributed to four ground users via four simultaneous downlinks. Although not a single ground-based connection was established, the network is fully connected and extends over an area of $\sim$1000 km.}
\end{figure} 

Especially for so-called type-0 phase matching~\cite{anwar2021entangled}, the spectral width of the photon pairs is relatively large ($\sim$100nm), which enables many wavelength channels.
In all spectral considerations, the transparency windows of the earth's atmosphere must be taken into account. 
Both transparency windows in the near infrared at around 800 nm and 1600 nm are broader than 100 nm. 
Therefore, wavelength channel numbers in the order of 1000 are possible in theory, which would lead to a 1000-fold increase in the key rate compared to a non-multiplexed QKD system.
Naturally, high channel numbers require a scalable approach to single-photon detection, such as imaging the channels on different areas on large SNSPD  arrays\cite{Rosenberg2013,Allman2015,Miki2014}.

For satellite-based QKD in a dual-downlink configuration, the de-multiplexing is implemented at the ground stations (see Figure~\ref{fig:multiplexing}). 
The only requirement for the polarization-entanglement source on the satellite is a broad spectrum and a narrowband pump laser for strong frequency anti-correlations.
Even the Micius satellite, with a few-nm spectral bandwidth can be used for wavelength-multiplexed QKD, since the optical ground stations can be retrofitted for frequency-resolved measurements.
The biggest obstacle in de-multiplexing of free-space links are angle-of-arrival (AOA) fluctuations caused by a turbulent atmosphere. 
AOA fluctuations lead to a spread of the incident angles on the dispersive devices used to frequency-resolve the spectrum, which can result in cross-talk between neighbouring frequency channels. 

Another intriguing extension of wavelength-multiplexed QKD are fully-connected QKD networks. 
This idea was brought forward in Ref.~\citenum{Wengerowsky2018}, where a single source of polarization-entangled photon pairs is frequency-demultiplexed and selectively multiplexd into four optical fibers, which allow four recipients to be fully connected in a network-topology sense.
Each of the recipients shares one anti-correlated frequency channel with all other users, which can be used to share entanglement with everyone in the network. 
The main benefit of this scheme is that each user only requires a single optical fiber, while all connections between the users are realized via frequency channels in a fully passive way.
In general, the users don't even need a frequency-resolved measurement, since the different frequency channels, corresponding to different users, can be identified by the arrival time difference of the photons.

Satellite-based QKD can exploit the idea of fully-connected networks in two different configurations (see Figure~\ref{fig:network_ground}).
The most straight-forward implementation of a network consists of a dual downlink from a satellite, where each ground receiver de-multiplexes the photons in wavelength channels.
After de-multiplexing, the channels are selectively multiplexed into several single-mode fibers and then guided to the end-users via optical fiber connections, e.g., within a city  (see Figure~\hyperref[fig:network_ground]{\ref*{fig:network_ground}a}). 
Crucially, all users between two remote cities are fully connected and can therefore simultaneously exchange cryptographic keys, a network architecture known as wide area network (WAN), while key exchange between local users can be established via ground-based wavelength-multiplexed fiber networks.

Another implementation of a quantum network based on frequency anti-correlations is based on a fully-connected network source on the satellite. 
The source itself is not necessarily much more complex than a common polarization-entanglement source, since the wavelength de-multiplexing and multiplexing on the satellite is accomplished by standard telecommunication components, such as WDMs, which are plug-and-play devices.
However, the source provides, e.g., four, separate output channels (see Figure~\hyperref[fig:network_ground]{\ref*{fig:network_ground}b}), which means that four simultaneous downlinks can be established to four different ground stations. 
Although this is indeed technically demanding regarding pointing and tracking, the prospect of a fully-connected quantum network over thousands of km is very intriguing, especially, since not a single terrestrial connection is involved in the network. 
While all ground-based users could also be provided with keys from successive flybys of a LEO satellite via dual downlinks, the very limited link time can be used in a much more efficient way by establishing multiple simultaneous downlinks. 

\subsection{High-dimensional QKD and spatio-temporal properties of photons} 
Most established QKD protocols~\cite{Bennett2014,Bennett1992} make use of two orthogonal quantum states, which lead to the measurement outcomes ``0'' and ``1'' and directly result in a binary secret key. 
Instead of only two states, e.g. the two polarization states of a photon, high-dimensional quantum states feature superpositions of many orthogonal quantum states, e.g. several energy levels of an atom or a photon.

The usefulness of high-dimensional quantum states for QKD has been investigated in the last decades\cite{BechmannPasquinucci2000,BechmannPasquinucci2000b, Cerf2002, Nikolopoulos2005, Sheridan2010}. 
Indeed, it turns out that high-dimensional QKD protocols are not only secure, but they are superior compared to two-dimensional protocols, since they exhibit an increased channel capacity\cite{Sheridan2010} and an increased resilience to noise~\cite{Ecker2019}. 
Instead of carrying only one bit of information, each copy of a high-dimensional quantum state potentially carries log\textsubscript{2}(d)$>$1 bits of information when it is measured, where d is the number of dimensions, of the state.
This is particularly relevant for QKD in high-loss scenarios such as downlinks from a satellite, where this higher information capacity can partially compensate for the unavoidable optical loss. 
It is therefore a direct measure to increase the key rate in QKD. 
High-dimensional states are also known to be more resilient to noise, which has been demonstrated for high-dimensionally entangled states\cite{Ecker2019}. 
For high-dimensional QKD protocols this implies that more noise can be tolerated before the key exchange must be aborted. 
An obvious contender for high-dimensional QKD protocols is satellite-based QKD, since theses protocols enable secret key exchange in the presence of stray light from the sun, artificial light sources and in the presence of imperfect sources, free-space channels and detection modules. 

Quantum states of light are an ideal platform for high-dimensional QKD, since they offer an infinite-dimensional state space which can be discretized in various ways. 
SPDC is particularly suitable for producing entanglement in spatiotemporal properties of photons, since the process is energy- and momentum conserving. 
Due to these conserved properties, high-dimensional entanglement in transverse spatial modes, momentum modes, frequency modes and temporal modes can be produced with modest experimental effort.
While all of these photonic properties can be harnessed in a laboratory environment, not all of them are suitable for free-space propagation in turbulent atmosphere.
Particularly, encodings based on spatial modes of light, such as orbital angular momentum modes~\cite{Erhard2017}, are degraded by wavefront distortions. 
Additionally, the divergence of higher spatial modes of light is greater than the divergence of the fundamental, Gaussian, mode, which necessitates larger receiving telescopes. 

The remaining candidates for free-space distribution of high-dimensional states are therefore the frequency and the temporal domain of photons. 
Both of them are well-suited for free-space propagation, since the atmosphere is neither dispersive nor varying at relevant timescales~\cite{Steinlechner2017}.
While the temporal and frequency properties are not disturbed in free-space propagation, most devices manipulating and measuring these properties rely on the photons occupying a single spatial mode, such as the fundamental mode guided in a single-mode fiber. 
Since light which leaves the transmitter optics in a single spatial mode is scrambled into a multimode beam by the turbulent atmosphere, strategies of either reducing the number of modes via adaptive optics or dealing with multimode light must be developed. 
The latter approach has been tackled for receiver interferometers by either introducing analyzers which are not sensitive to small AOA fluctuations\cite{Steinlechner2017}, or by imaging techniques which reverse parts of the beam propagation or provide a field-widened angular tolerance~\cite{Cahall2020,Jin2018, Jin2019,Bulla2022}

\begin{figure} [t]
\begin{center}
\includegraphics[width=0.6\textwidth]{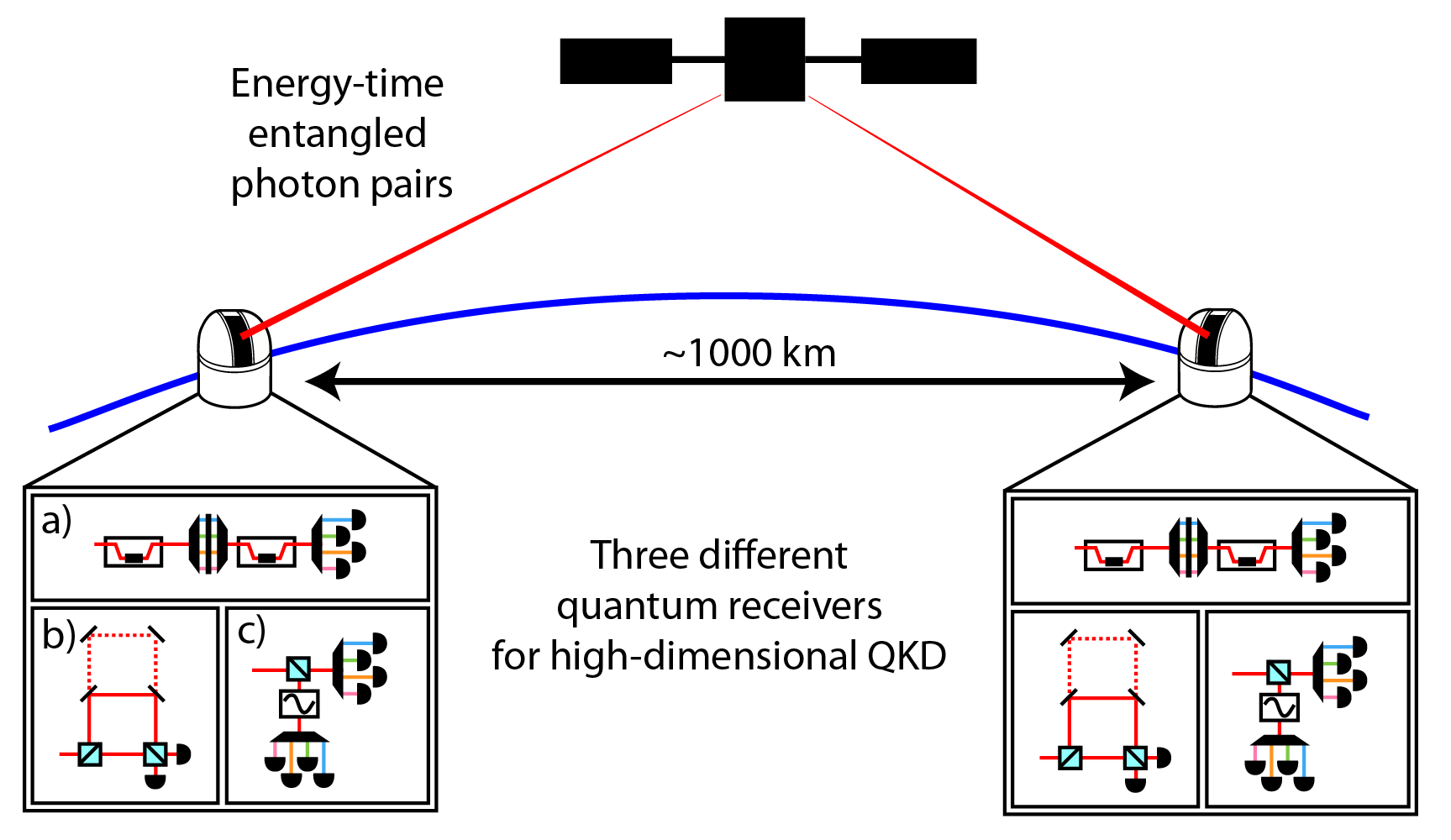}
\end{center}
\caption[example] 
{\label{fig:highdim} 
High-dimensional QKD via satellite downlinks. Pairs of photons entangled in the energy-time domain are produced in a satellite and distributed to the communicating parties on ground. Quantum receiver for a) the frequency domain using electro-optic modulators and spectrometers, b) the time domain using imbalanced interferometers and c) the time-frequency domain using spectral and temporal resolution are illustrated. The main benefit of this approach is a reduced complexity of the quantum payload, while the quantum receivers on ground can be easily upgraded.}
\end{figure}

QKD protocols can be implemented in various ways in the energy-time domain of photons. 
All of the protocols either operate entirely on the frequency or the time domain, or simultaneously on both domains in time-frequency protocols~\cite{Islam2017,Rdiger2017,Zhang2014,Nunn2013,Mower2013}. 
Both domains can be conveniently divided into time- or frequency bins, while discretizations into more complex modes have been demonstrated\cite{Brecht2015,Raymer2020}.
In SPDC, time-bins can be produced in two ways. 
They can either be created by a pulsed pump laser \cite{Tittel2000,deRiedmatten2004}, or by discretizing continuous temporal superpositions created by a continuous-wave pump laser into arbitrary subspaces~\cite{Kwiat1993}.
Two convenient measurement bases in the time domain are the time-of-arrival basis and a superposition basis which can for example be implemented by an imbalanced interferometer~\cite{Ecker2021b}, as illustrated in Figure~\hyperref[fig:highdim]{\ref*{fig:highdim}b}.

While the SPDC process creates frequency entanglement, a discretization into frequency bins is most commonly achieved by a cavity ~\cite{Xie2015}, by spatial light modulators \cite{Bessire2014} or by Hong-Ou-Mandel interference \cite{Chen2021}. 
Superposition measurements of frequency-bins involving a single photon are hard to perform and have only seen few implementations using electro-optic modulators and pulse shapers\cite{Lukens2016}, as illustrated in Figure~\hyperref[fig:highdim]{\ref*{fig:highdim}a}. 
Therefore, hybrid approaches are advantageous, where both the temporal and the frequency domain are addressed\cite{Islam2017,Rdiger2017,Zhang2014,Nunn2013,Mower2013}, and no superposition basis in either domain must be accessed, as illustrated in Figure~\hyperref[fig:highdim]{\ref*{fig:highdim}c}. 
These time-frequency protocols utilize the incompatibility of simultaneously obtaining information about the state in two conjugate domains, such as energy and time. 

In large-alphabet QKD protocols, the time domain is divided into time-frames, where each frame consists of a fixed number of time-bins~\cite{AliKhan2007,Zhong2015}. 
A photon detection in a certain time-bin corresponds to a symbol of the QKD alphabet, which yields several secret key bits. 
The security of the protocol is guaranteed by measurements in a superposition basis.
A recently developed method simultaneously uses multiple subspaces of high-dimensional states and can thus overcome high levels of noise \cite{Doda2021}, which was already experimentally demonstrated in the temporal domain over a free-space link~\cite{Bulla2022} and for path encodings in the laboratory \cite{Hu2021}.

High-dimensional states almost always result in higher key rates and increased resilience to noise. 
However, due to imperfections of the measurement devices, the maximal key rate can nevertheless be achieved for lower-dimensional states. 
Especially for non-static quantum links, this optimal dimensionality can vary drastically, which necessitates a dimension-adaptive approach to QKD~\cite{Ecker2019}. 
Additionally, the QKD protocol itself can be switched depending on the quantum link conditions \cite{Bouchard2018}.
Advanced QKD systems will thus benefit from maximizing the key rate by dynamically optimizing the parameter space consisting of the quantum state dimension, the protocol type and the source brightness.
An alternative approach towards high-dimensional QKD is the simultaneous utilization of several properties of a single photon \cite{chapman19, Vergyris2019}. 
Photon pairs produced in SPDC are simultaneously entangled in several properties, which results in so-called hyperentanglement\cite{Kwiat1997}. 
An experiment over a 1.2 km-long free-space demonstrated the transmission of hyperentangled states through turbulent atmosphere and showed that high-dimensional entanglement can be certified with high fidelities~\cite{Steinlechner2017}. 

High-dimensional QKD is an important extension of conventional, two-dimensional QKD, since it simultaneously increases the key rate and the noise robustness. 
It is important to note that the source complexity is not necessarily increased for the creation of high-dimensional states.
On the contrary, photon pairs generated in SPDC are intrinsically high-dimensional entangled in the frequency and temporal domain and do not require overlap of two SPDC processes as is the case for polarization-entangled photon pair sources.
This is very attractive for space-borne sources, since none or only minor upgrades of the source are required, while the intricacies of coping with high-dimensional QKD is shifted to the quantum receivers, which can be upgraded in retrospect.  
An additional challenge in satellite-based implementations of high-dimensional QKD is the propagation of the photons through turbulent atmosphere. 
The resulting AOA fluctuations and spatial mode scrambling must be factored into the design of the receiver modules.

\subsection{Adaptive optics for quantum communication}
\label{sec:adaptopt}
After propagating through the atmosphere, the wavefront of optical beams is distorted due to time-varying density fluctuations of turbulent air. 
Adaptive optics (AO) systems consist of a wavefront sensor and a deformable mirror, which can correct wavefront distortions on relevant timescales by updating the deformable mirror in the kHz regime.
While AO has a long history in ground-based astronomy, it becomes increasingly relevant for free-space optical communication~\cite{Weyrauch2002,Fischer2017}.
The compensation scheme in optical feeder links is different depending on the link configuration. 
For uplinks, the communication link is pre-compensated, while for downlinks the communication link is post-compensated. 

The benefits of using AO systems for quantum communication are similar to the benefits of using AO systems for optical communication. 
Beam spreading leads to an effective diameter of the free-space beam which is typically much larger than the telescope aperture. 
By employing AO systems, beam spreading is reduced, which leads to a reduction of geometrical loss. 
A major benefit of using AO systems for quantum communication is the ability to launch the multi-mode free-space beam into single-mode fibers. 
Quantum receivers are coherent detectors which greatly benefit from single-mode operation. 
Additionally, single-mode coupled light is compatible with fiber networks.
Metropolitan fiber networks are essential for the deployment of secure keys, since they distribute the photons from the optical ground stations to the end users.
AO systems therefore constitute an important interface between free-space and fiber-based quantum networks.

\subsection{Space-based quantum repeaters} 
The distribution of photons via optical fibers is limited by absorption. 
This leads to an exponential dampening with increasing fiber length and limits the achievable transmission distance to a few hundred km~\cite{Wengerowsky2018,Neumann22}.
While classical information can be amplified without loss of information, which is essential for our fiber-optic telecommunication networks, this is not true for quantum information. 
The no-cloning theorem prohibits any attempts to copy quantum states, which renders a quantum repeater a fictional device at first sight.
However, the quantum repeater does not amplify quantum states, but instead increases the probability of a successful distribution of an entangled state between distant communicating parties by dividing the entire transmission distance into many elementary links~\cite{Briegel1998}. 
6In each of these elementary links, entanglement is distributed and stored in quantum memories (QM) upon successful distribution. 
The last and crucial step of the quantum repeater protocol consists of jointly measuring the entangled photons stored in the QM in such a way, that entanglement between the two most-distant communicating parties remains, a procedure known as entanglement swapping~\cite{ukowski1993}.

While recent laboratory experiments have successfully demonstrated an elementary quantum repeater node~\cite{Liu2021,LagoRivera2021}, the bottleneck is the storage of photons in QM.
Many research groups are working on different implementations of a QM ~\cite{Lvovsky2009,Sangouard2011}, but most realizations do not simultaneously fulfil the stringent requirements of a quantum repeater.
Even for the most optimistic estimates of a future QM, the maximal achievable distance which can be bridged by a fiber-based quantum repeater is around 2000 km before the secret key rate drops to impractical levels~\cite{Liorni2021}.
Therefore, QKD on a global scale can only be realized by combining satellite-assisted quantum communication and quantum repeaters.
The space-based quantum repeater indeed overcomes loss to an extent that allows distribution of secret keys on intercontinental distances at practical rates~\cite{Liorni2021}.

\begin{figure} [t]
\begin{center}
\includegraphics[width=0.6\textwidth]{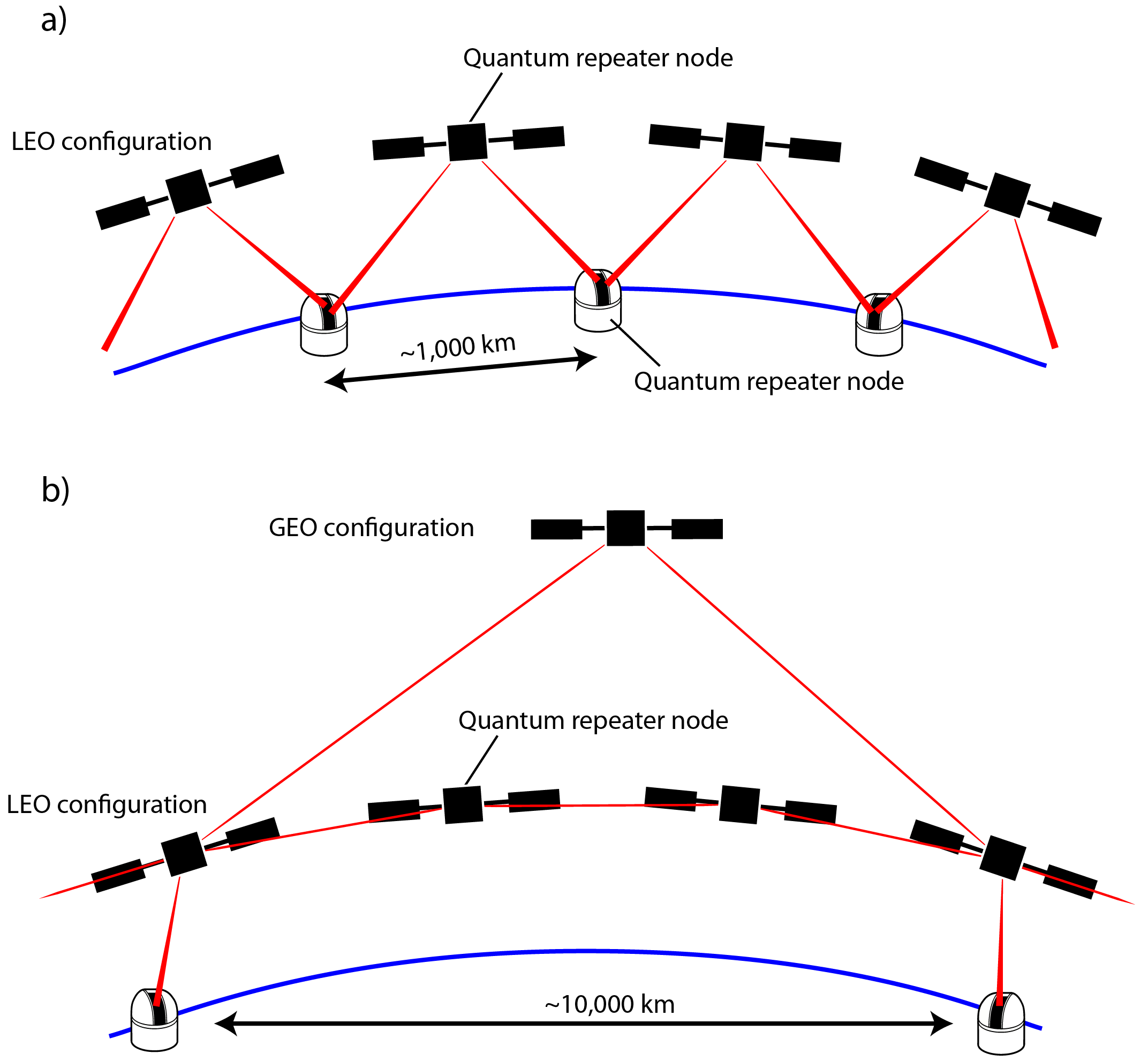}
\end{center}
\caption[example] 
{\label{fig:repeaterchain} 
The two most promising space-based quantum repeater configurations. a) Hybrid space-based quantum repeater~\cite{Boone2015}. Photon pair sources on satellites distribute entanglement between ground stations via dual downlinks. The photons are stored in QMs on ground and jointly measured after successful distribution. b) Fully space-based quantum repeater~\cite{Sangouard2011,Gndoan2021}. Both the photon pair sources and the QMs are installed on satellites. The quantum repeater protocol is entirely performed in space, and downlinks are only established if two communicating parties require shared entanglement.}
\end{figure} 

A first step towards space-based quantum repeater networks are single-node quantum repeaters, which can be utilized for so-called memory-assisted QKD (MA-QKD) protocols. 
In MA-QKD, the successfully distributed photons of a single down or uplink are stored in QM on a satellite~\cite{Luong2016,Panayi2014}.
Upon successful distribution of two independent photons, the two quantum memories on the satellite are read out and jointly measured, which concludes the protocol. 
The rate-distance scaling is favourably compared to a dual down-or uplink, since the two photons are independently transmitted. 
While the uplink scenario is more demanding and lossier, it is an implementation of measurement-device independent QKD\cite{Wang2021} and therefore avoids trust in the satellite.
The secure key rates achievable in MA-QKD with multimode quantum memories are roughly an order of magnitude higher than what is achievable with entanglement-based QKD.

Another single-node quantum repeater consists of a single LEO satellite containing an entangled photon pair source and a QM payload.
Here, a single satellite can be used to distribute secret keys on a global scale \cite{Wittig2017}.
One photon of the photon pair source is stored in the QM while the other photon is distributed to a ground station via a downlink. 
The satellite then follows its orbital trajectory and releases the stored photon at a later time once another ground station is in line of sight. 
While this scheme could replace a whole quantum repeater network, the effective rates are low, since the bottleneck of the scheme is the orbital velocity of the LEO satellite. 
Additionally, the demands on the storage time of the QM are extremely high, since the quantum states must be stored until the second downlink is established, which can take several hours. Nevertheless, the scheme is an elegant solution for global entanglement distribution and it is very resource efficient since it requires only a single satellite.

The so-called hybrid space-based quantum repeater~\cite{Boone2015} is not entirely space-based, since the satellites are merely equipped with entangled photon pair sources, while the quantum repeater nodes on ground accommodate the QMs (see Figure ~\hyperref[fig:repeaterchain]{\ref*{fig:repeaterchain}a}). 
If all QMs in the ground stations are loaded with photons from satellite dual downlinks, the entanglement can be swapped. 
The obvious benefit of this approach is that the demanding parts of the quantum repeater are on ground where they can be further developed, while the distribution of entangled photon pairs from satellites has already been successfully demonstrated \cite{Yin2017}. 
Although this scheme is relatively lossy due to several downlinks, it beats the direct transmission of photons from high-orbit satellites. 
Even GEO satellites, although they can in principle operate continuously in contrast to LEO quantum repeaters, are impractical beyond $\sim$10,000 km ground distance~\cite{Boone2015}.

The fully space-based quantum repeater does not depend on ground stations as quantum repeater nodes \cite{Sangouard2011,Gndoan2021}. 
While ground-based quantum repeater nodes are advantageous since they are accessible, the required downlinks add unnecessary loss to the link budget. 
In the fully space-based quantum repeater, the ground-based repeater nodes are replaced by satellite-based repeater nodes (see Figure ~\hyperref[fig:repeaterchain]{\ref*{fig:repeaterchain}b}).
This is highly advantageous because the satellite-based nodes receive diffraction-limited photons without wavefront disturbances and can therefore be directly coupled into single-mode fibers for interfacing with QMs.
Several repeater stations can be chained to form a quantum repeater network with global coverage. 
This quantum repeater configuration is by far the most efficient solution for global entanglement distribution.

Space-based quantum repeater infrastructures are currently the only way to achieve truly global quantum communication. 
The bottleneck for their implementation is currently the state-of-art of QMs. 
Their storage time is of utmost importance, since the communication distances are vast and the losses are relatively high, which requires storage times in the order of seconds.
For the same reason, QMs must simultaneously hold many photons in a multi-mode memory, since otherwise the secret key rates will always be limited by the specifications of QMs.

An alternative approach to extending the transmission distance of free-space links is to bypass the diffraction limit in a clever way. 
While the divergence of a free-space beam constitutes a fundamental limit on the size of the beam spot at a certain distance from the transmitter telescope, a system of telescopes or lenses can refocus a beam indefinitely. 
A satellite-based or drone-based~\cite{Liu2021b} system of relay telescopes can therefore extend the transmission distance substantially without relying on quantum memories or a quantum repeater.  

Another active area of research is the architecture of a space-based world-spanning quantum network, which raises questions such as continuous coverage, possible constellations, network connectivity and network topologies \cite{Khatri2021, Wallnfer2022,Harney2022,Wei2022}.
Entanglement is an irreplaceable resource in quantum information processing, which is essential far beyond quantum cryptography. Among others, quantum entanglement enables distributed quantum computing, long-baseline quantum sensing \cite{Zhang2021}, quantum state transfer via quantum teleportation \cite{Pirandola2015} and fundamental physics experiments \cite{Mohageg2021}. The utilization of entanglement in a quantum repeater as a means to overcome long distances in quantum communication is therefore an ideal match.

\section{CONCLUSION AND DISCUSSION}
\label{sec:conclusion}
Space-based QKD is currently the only viable option to overcome the huge optical losses in terrestrial distribution of photons.
The feasibility of establishing secure keys over long distances by employing optical links between satellites and ground stations has been successfully demonstrated.
The remaining challenge is to advance space-based QKD to the point, where the key rates enable commercially successful operation of quantum space links.  
In this article, we reviewed some of the most promising methods for increasing the key rate and decreasing the system costs in space-based QKD. 
Naturally, the ease of integration into existing infrastructure and the timeline for establishing new infrastructure varies drastically between these approaches.  

The most technologically immature approach is the establishment of a space-based quantum repeater. 
While the quantum repeater is currently the only means of distributing entanglement over global distances, the technological readiness level is relatively low, since the basic building blocks have only just been demonstrated in laboratory environments. 
The deployment of bright integrated photon-pair sources can increase the key rate under certain circumstances and decrease size, weight and power consumption drastically. 
Since the photon-pair sources in entanglement-based QKD are spaceborne, the remaining steps from a breadboard model in the laboratory towards a flight model are space environment readiness and qualification.
Adaptive optics already has been employed for optical space links and is vital for interfacing fiber-based quantum networks with satellite-based quantum networks. 
However, adaptive optic systems dramatically increase the costs and complexity of the ground receiver and suitable passive alternatives for coupling light into optical fiber are required for wide-spread deployment of ground terminals.
Multiplexed QKD and high-dimensional QKD in the frequency or time domain are currently the most promising candidates for increasing the key rate in satellite-based QKD.
Not only do they promise substantial increases in key rates, but they also mainly require changes to the ground receivers.
For every change in the ground receivers in prepare-and-measure QKD, also the transmitter in the space segment must be changed. 
In our entanglement-based scenario, the space segment is much more versatile and a single photon pair source in the space segment can serve multiple purposes and protocols.
Wavelength-division multiplexing requires frequency-resolved detection of single photons, while high-dimensional quantum receivers additionally require superposition measurements in frequency or time.  
First demonstrations of such measurements after terrestrial free-space links have shown the feasibility of this approach. 

The timeline of space missions, from an idea to the launch pad, is notoriously long.
Even more so, since the availability of components for security systems with national or regional independence can mark a critical path for development. 
It is therefore essential to collect potential candidate technologies for future QKD missions as early as possible in order to start an informed discussion.  
Since Europe's space industry must first catch up with other players when it comes to satellite-based QKD, every scientific and technological advancement should be embraced and exploited. 
In this context, it is important to again highlight that entanglement-based QKD lays the foundation for the development towards a global quantum network.
Therefore, advances in entanglement-based QKD in space will not only increase the capabilities of quantum cryptography but will also stimulate the development of the quantum internet.

\acknowledgments
We gratefully received funding from ESA European Space Agency Contract 4000134348/21/NL/GLC/ov.


\end{document}